\documentclass[
 reprint,
superscriptaddress,
 amsmath,amssymb,
 aps,
superscriptaddress,
 amsmath,amssymb,
 aps,longbibliography
]{revtex4-1}

\usepackage{array}
\usepackage{booktabs}
\usepackage{graphicx}
\usepackage{dcolumn}
\usepackage{bm}
\usepackage{amsmath,amssymb,amstext,amsthm}
\usepackage{braket}
\usepackage{bbold}
\usepackage{bbm}
\usepackage{xcolor}
\usepackage{ gensymb }
\usepackage{soul}
\setstcolor{red}
\usepackage{ mathrsfs }
\usepackage{cancel}
\usepackage{array}

\usepackage{hyperref}

\hypersetup{colorlinks,linkcolor=blue,citecolor=blue,urlcolor=blue}

\DeclareMathOperator{\arcsinh}{arcsinh}
\usepackage{rotating}  

\newcommand{\Pend}{pendell\"osung }

\usepackage{etoolbox}

\AtBeginEnvironment{equation}{\vspace{6pt}}

\begin{document}

\preprint{APS/123-QED}
\setlength{\abovedisplayskip}{1pt}
\title{A unified quantum random walk model for internal crystal effects in dynamical diffraction}

\author{Owen Lailey} 
\email{oalailey@uwaterloo.ca}
\affiliation{Institute for Quantum Computing, University of Waterloo,  Waterloo, ON, Canada, N2L3G1}
\affiliation{Department of Physics and Astronomy, University of Waterloo, Waterloo, ON, Canada, N2L3G1}

\author{Dusan Sarenac}
\affiliation{Department of Physics, University at Buffalo, State University of New York, Buffalo, New York 14260, USA}

\author{David G. Cory}
\affiliation{Institute for Quantum Computing, University of Waterloo,  Waterloo, ON, Canada, N2L3G1}
\affiliation{Department of Chemistry, University of Waterloo, Waterloo, ON, Canada, N2L3G1}

\author{Michael G. Huber}
\affiliation{National Institute of Standards and Technology, Gaithersburg, Maryland 20899, USA}

\author{Dmitry A. Pushin}
\email{dmitry.pushin@uwaterloo.ca}
\affiliation{Institute for Quantum Computing, University of Waterloo,  Waterloo, ON, Canada, N2L3G1}
\affiliation{Department of Physics and Astronomy, University of Waterloo, Waterloo, ON, Canada, N2L3G1}

\date{\today}

\pacs{Valid PACS appear here}

\begin{abstract}
    The theory of dynamical diffraction (DD) in perfect crystals is the backbone of high-precision neutron and X-ray diffraction experiments, enabling accurate determination of crystal structure factors and the realization of perfect crystal interferometers. In practice, however, real crystals exhibit deformations and imperfections, including surface roughness, defects, temperature gradients, angled crystal faces, and curvature, that degrade interferometer performance and are difficult to model using conventional DD theory, particularly in complex geometries. To address these challenges, a quantum information (QI) model for DD has been under development, with demonstrated experimental agreement for both ideal crystals and in the presence of some imperfections such as surface roughness and defects. Here, we present a unified quantum random walk model that is now suitable for reproducing all established DD effects. We demonstrate this by incorporating a broad range of internal crystal effects influencing DD intensity distributions, including linear temperature gradients, the DD Talbot effect, and angled or miscut crystals. These results establish the QI model as a comprehensive and flexible framework for experimental analysis, as well as for the design of next-generation perfect crystal neutron interferometers and neutron optical components, such as condensing monochromators. 
\end{abstract}
\maketitle

\section{Introduction}
Bragg diffraction plays a fundamental role in neutron and X-ray experiments, particularly in applications involving perfect crystal interferometers, which have been integral to research across multiple fields for over six decades~\cite{bonse_first, xray_hist, sam_book, sears}. These interferometers have enabled significant advancements in the development of phase-contrast imaging~\cite{davis1995phase, momose1996phase, pushin2007reciprocal}, high-precision lattice constant measurements for redefining the kilogram~\cite{lattice_constant, ferroglio2008si, massa2011measurement}, and in the search for fifth forces~\cite{li2016neutron, heacock2021pendellosung}.

Achieving high-contrast perfect crystal interferometers is challenging due to the rigorous demands they place on fabrication tolerances and environmental control~\cite{arif1994x, huber2024achieving}. For instance, crystal blade lattice planes must be aligned to within a few angstroms~\cite{Rauch_2012, massa2009observation, heacock2017neutron}, and angular deviations from the Bragg condition must remain within the narrow constraints of the Darwin width, typically limited to just tens of microradians~\cite{sam_book, authier2006dynamical}. Furthermore, the crystal must be effectively isolated from external disturbances, as even low frequency vibrations or temperature fluctuations on the order of a few millikelvin will reduce contrast~\cite{pushin2009decoherence, pushin2011experimental, saggu2016decoupling}.

Advancements in reproducible, high-contrast perfect crystal interferometer fabrication methods~\cite{heacock2018increased, heacock2019measurement, huber2024achieving} could enable novel high-precision experiments, such as tests of gravity modifications~\cite{grav_interferometry}, large-area phase measurements~\cite{lemmel2022neutron}, and neutron magnetic \& electric dipole moment measurements~\cite{zeilinger1984symmetry, dombeck2001measurement, fedorov2010measurement, neutron_reflex, gentile, lailey2026perfect}. This underscores a need to improve our understanding of experimental imperfections so that we can characterize and mitigate them.

To achieve this, a quantum information (QI) model for dynamical diffraction (DD) has been under development, which simulates both neutron and X-ray propagation through perfect crystals as a quantum random walk through a lattice of nodes~\cite{Nsofini_2016, nsofini2017noise, nsofini2019coherence, nahman2022generalizing, neutron_cav, lailey2026perfect}. To establish a comprehensive simulation framework for modeling high-precision experiments, it is essential to incorporate the effects of crystal imperfections, as well as strain and stress. Initial efforts in this direction applied the QI model to a neutron cavity formed by two Bragg blades, introducing a toolbox for modeling crystals with rough surfaces, which is often impractical using conventional DD~\cite{neutron_cav}.

In this work, we present a unified quantum random walk framework for both neutron and X-ray Bragg diffraction experiments, suitable for reproducing all established DD effects within a single model. We demonstrate this capability by incorporating a broad range of internal crystal effects and deformations. We analyze elastically deformed crystals subject to a uniform temperature gradient and compare the results with experimental data~\cite{hart}. We also consider angled crystal faces, which may arise from fabrication errors or be intentionally introduced~\cite{fankuchen1937condensing}, and demonstrate how crystal asymmetry strongly modifies the diffracted intensity and beam divergence. Finally, we investigate the DD analogue of the Talbot effect, which significantly modifies neutron/X-ray wavefunction self-interference within the crystal~\cite{balyan2019a, balyan2019b, balyan2019c, balyan2020spherical, balyan2020x, balyan2021x}. To understand this effect, we introduce for the first time how coherent plane-wave DD results can be implemented within the QI model, in contrast to previous spherical-wave approaches~\cite{Nsofini_2016, nsofini2017noise, nsofini2019coherence, nahman2022generalizing, neutron_cav, lailey2026perfect}. 

\section{Theoretical Framework}
\subsection{\Pend oscillations}
In the Laue geometry of DD theory, defined by Bragg planes perpendicular to the crystal surface, an incident beam of neutrons or X-rays enters one face of the crystal and exits through the opposite, with both the transmitted (O-Beam) and Bragg-diffracted beams (H-Beam) propagating in the forward direction. Unlike the simpler kinematic approximation, which assumes single scattering, DD theory accounts for multiple scattering events within the crystal volume and the coupling between incident and diffracted wavefields due to the periodic crystal potential. Under the two-wave approximation, where only one set of Bragg planes contributes significantly to the scattering, the neutron or X-ray wave equation within the crystal admits two Bloch wave solutions. Each Bloch wave is a coherent superposition of a transmitted and a diffracted component, and together they form the complete solution for the internal wavefield. Thus, four plane waves propagate inside the crystal (two transmitted and two diffracted), grouped into two Bloch wave eigenstates with distinct wavevectors ($k^\alpha$ and $k^\beta$)~\cite{authier2006dynamical, sam_book}.

As these two Bloch waves propagate through the crystal, their differing longitudinal wavevector components lead to a periodic modulation of the relative phase between them. This produces an interference effect observable as a spatial redistribution of intensity between the transmitted and diffracted beams, known as \Pend oscillations. These oscillations occur as a function of crystal thickness $t$ or incident wavelength $\lambda$, and are a direct consequence of the coherent superposition of the two Bloch waves~\cite{shull1968observation}. The \Pend period is given as:
\begin{equation}
  \Delta_H = \frac{2\pi}{|{k^\alpha - k^\beta}|} = \frac{1}{C}\frac{\pi V \cos(\theta_B)}{\lambda|F_H|},
  \label{deltah}
\end{equation}
where $V$ is the unit cell volume, $\theta_B$ is the Bragg angle, $\lambda$ is the wavelength, $C$ is the polarization factor (unity for neutrons, polarization-dependent for X-rays), and $F_H$ is the structure factor. The nature of $F_H$ depends on the scattering mechanism: for X-rays it is related to the electron density, while for neutrons it depends on the nuclear scattering length density~\cite{sears, sam_book}.

\subsection{Quantum information model}
\label{qi_sec}
The QI model describes neutron and X-ray propagation through perfect crystals as a quantum random walk through a two-dimensional lattice of nodes~\cite{Nsofini_2016, nsofini2017noise, nsofini2019coherence, nahman2022generalizing, neutron_cav, lailey2026perfect}. The beam input state to a single node (indexed with $\eta$) is given by the two-level system state vector:
\begin{equation}
    \psi_\eta = \alpha_\eta \ket{a_\eta }+ \beta_\eta \ket{b_\eta }
    \quad\mathrm{or}\quad 
    \begin{pmatrix}
           \alpha_\eta \\
           \beta_\eta \\
    \end{pmatrix},
    \label{input}
\end{equation}
where $\ket{a_\eta}$ and $\ket{b_\eta}$ are the states propagating upwards or downwards respectively into the $\eta^{th}$ node. For a column of $h$ nodes, the input beam wavefunction is represented as:
\begin{equation}
    \psi = \begin{pmatrix}
        \vdots \\ \alpha_\eta \\ \beta_\eta  \\ \vdots \\ \alpha_{h} \\ \beta_{h}
    \end{pmatrix},
\end{equation}
whereby the nodes act as unitary operators on the input wave function resulting in transmission and reflection to the neighbouring nodes. The unitary operator for a node is given by:
\begin{equation}
    U_\eta = \begin{pmatrix}
        \cos\gamma & \sin\gamma \\
        -\sin\gamma & \cos\gamma  \\
    \end{pmatrix},
    \label{U}
\end{equation}
where $\gamma$ controls the transmission and reflection amplitude. There is an equivalence relation between simulation and experimental parameters given by: 
\begin{equation}
    N \cdot \gamma = \frac{\pi t}{\Delta_H},
    \label{eq:equivalence}
\end{equation}
where $N$ is the number of nodes in simulation space to simulate the corresponding crystal thickness $t$ for \Pend length $\Delta_H$ from Eq.~\ref{deltah}. 

Throughout prior QI model developments~\cite{Nsofini_2016, nsofini2017noise, nsofini2019coherence, nahman2022generalizing, neutron_cav, lailey2026perfect}, the focus has been on spherical-wave formulations of DD, consistent with the Takagi–Taupin equations originally developed to analyze strain in X-ray diffraction experiments~\cite{sam_book}. While this approach naturally captures real-space intensity profiles of diffracted beams, it does not directly reveal the angular-spectrum structure underlying many fundamental DD phenomena. In particular, effects that depend explicitly on well-defined plane-wave components, such as misset-angle dependence (rocking curves) and the DD Talbot effect, are most transparently described in the plane-wave limit. Motivated by this, here we develop the coherent plane-wave formulation within the QI model, completing its exact connection to foundational DD results~\cite{sam_book}, and enabling exploration of the internal crystal Talbot effect in Sec.~\ref{talbot_section}. 

Consider a neutron or X-ray plane wave with wavevector magnitude $k_0 = 2\pi/\lambda$ incident on a Laue crystal, with the reciprocal lattice vector oriented along the $x$ axis. For a small misset angle $\delta\theta$ relative to the Bragg angle $\theta_B$, the wavevector component along the crystal surface is $k_x = k_0 \sin(\theta_B + \delta\theta)$. The physical spatial coordinate along the crystal surface $x$ takes on discrete values in the simulation space corresponding to the node positions. The node spacing is $d_{\mathrm{nodes}} = \gamma\Delta_H/\pi$, from setting $N=1$ in Eq.~\ref{eq:equivalence}. Notice that for small $\gamma$, $d_{\mathrm{nodes}}$ can approach the true physical lattice spacing of the crystal $d_{\mathrm{crystal}}$. Altogether, the physical coordinate of a node along the crystal surface is given by $x = 2\eta d_{\mathrm{nodes}}$, where $\eta \in \mathbb{Z}$ is the node index. Subbing in the expressions for $k_x$ and $x$, and approximating for small $\delta\theta$, we obtain
\begin{equation}
    \psi_{\mathrm{plane}} = A_0 \exp(i k_x x) \sim A_0 \exp\left(i \frac{2\pi \eta\, \delta\theta}{\tan(\theta_B)} \frac{d_{\mathrm{nodes}}}{d_{\mathrm{crystal}}} \right),
    \label{plane}
\end{equation}
where $A_0$ is the incident wave amplitude. This expression represents a plane wave incident on the discretized crystal, with a phase gradient determined by the angular deviation ($\delta \theta$) from the Bragg condition.

\begin{figure}
    \centering
    \includegraphics[width=\linewidth]{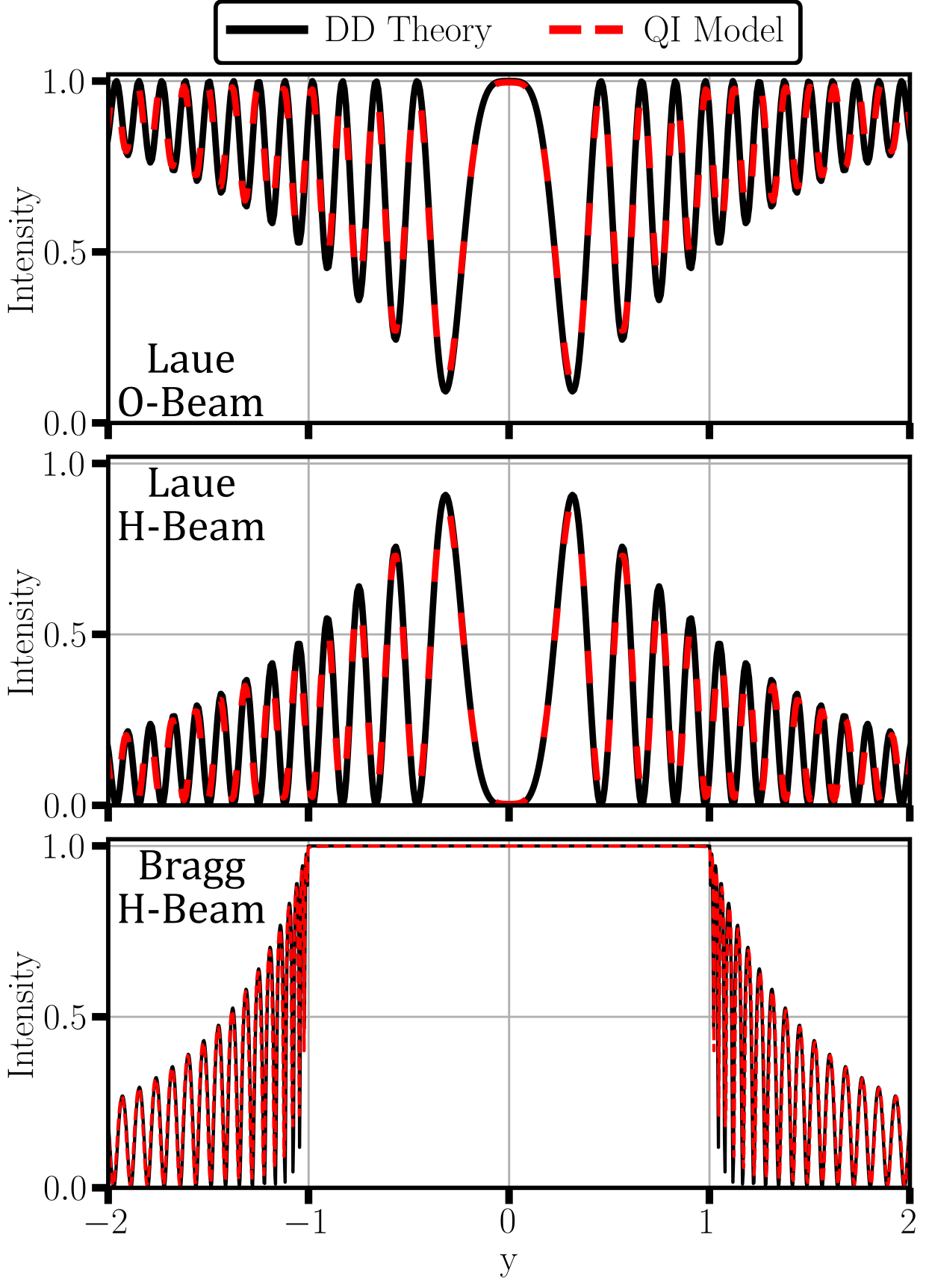}
    \caption{Transmitted (O-beam) and diffracted (H-Beam) intensity profiles in both the Laue and Bragg geometry for an incident misset monochromatic plane wave. The misset angle $\delta\theta$ is related to the conventional dimensionless misset parameter $y$ according to $y = 2\Delta_H \sin(\theta_B)\delta\theta/\lambda$~\cite{sam_book, sears}. The crystal thickness is $t = 10\Delta_H$ (220), $\lambda = 2.72$~\AA, $\gamma = \pi/52$, and $0\leq\eta\leq4000$. The QI model and DD equations (Eq.~\ref{DD_laue},\ref{DD_bragg}) are in excellent agreement.}
    \label{fig:plane}
\end{figure}

In Fig.~\ref{fig:plane}, we plot the QI model results for incident monochromatic plane waves within DD theory for both Laue and Bragg geometries for a crystal thickness of $t = 10\Delta_H$. The results are compared with the standard DD expressions given by:
\begin{equation}
    I = 
    \begin{cases} 
    y ^2\frac{\sin^2(\pi t \sqrt{1 + y^2})}{(1 + y^2)} + \cos^2(\pi t \sqrt{1 + y^2}), & \text{O-Beam}\\
    \frac{\sin^2(\pi t \sqrt{1 + y^2})}{(1 + y^2)}, & \text{H-Beam}
    \end{cases}
    \label{DD_laue}
\end{equation}
for the Laue geometry and
\begin{equation}
    I_H = 
    \begin{cases}
    \frac{
\sinh^2\!\left(\pi\, t\, \sqrt{1 - y^2}\right)
}{
(1 - y^2) + \sinh^2\!\left(\pi\, t\, \sqrt{1 - y^2}\right)
}, & |y| \leq 1 \\
    \frac{
\sin^2\!\left(\pi\, t\, \sqrt{y^2 - 1}\right)
}{
(y^2 - 1) + \sin^2\!\left(\pi\, t\, \sqrt{y^2 - 1}\right)
}, & |y| \geq 1
    \end{cases}
    \label{DD_bragg}
\end{equation}
for the Bragg reflected beam, with the conventional dimensionless misset parameter $y = 2\Delta_H \sin(\theta_B)\delta\theta/\lambda$~\cite{sam_book, sears}. Excellent agreement with theory is obtained: in the Bragg geometry, the characteristic unity reflectivity for $-1 < y < 1$ is recovered, while in the Laue geometry, the transmitted intensity reaches unity at $y = 0$. This formulation naturally extends to multiple-crystal configurations, such as interferometers.

This establishes the QI model as a valid, single framework for both plane-wave and spherical-wave regimes.

\section{Internal crystal effects, imperfections, and deformations}
The QI model is capable of simulating a wide range of internal crystal effects, imperfections, and deformations, as summarized in Table~\ref{table}. Previous work has addressed the effects of surface roughness and defects~\cite{neutron_cav, lailey2026perfect}. In the present study, we consider linear temperature gradients, the DD Talbot effect, and angled crystal geometries. We also discuss further implementations, such as the treatment of bent crystals and the incorporation of spin.

\subsection{Linear Temperature Gradients}
Crystals can be elastically deformed by a uniform temperature gradient applied normal to the Bragg planes, which induces a phase difference between the coherently interfering wavefields within the crystal and modifies the resulting \Pend oscillations, as demonstrated by Hart in Ref.~\cite{hart}.

\begin{table}[htbp]
\centering
\caption{Internal crystal effects, deformations, or imperfections and the proposed implementation with the QI model, with the listed equations from this work. The references highlight experimental demonstrations of the effects. Surface roughness and defects have been considered in previous work~\cite{neutron_cav, lailey2026perfect}; linear temperature gradients, the DD Talbot effect, and angled crystals are demonstrated in this work; bent crystals and incorporating spin are directions for future work.}
\label{table}
\vspace{0.1cm}
\begin{tabular}{lcr}
\toprule
\parbox{4cm}{\raggedright \textbf{Internal crystal effects, deformations, or imperfections}} & \parbox{3cm}{\textbf{QI Model Implementation}} & {\raggedleft \textbf{Refs.}} \\
\midrule
\hline \hline
Surface roughness + defects & Modify $\gamma$, Eq.~\ref{U}  & \cite{neutron_cav} \\
 \hline
Linear temperature gradient & Scale $N$, Eq.~\ref{eq:gamma} & ~\cite{hart} \\
Angled or miscut crystals & Shape crystal, Eq.~\ref{Ufree} & ~\cite{fankuchen1937condensing} \\
DD Talbot effect & $\psi_{plane}$, Eq.~\ref{plane} & ~\cite{balyan2019a, balyan2019b, balyan2019c, balyan2020spherical, balyan2020x, balyan2021x}\\
 \hline
Bent crystals & Phase gradients & ~\cite{klar1973dynamical, albertini1976simple, albertini1977dynamical} \\
\parbox{4cm}{\raggedright Magnetic fields + Spin-orbit (Schwinger)} & Introduce $U_{spin}$ & {\raggedleft \cite{gentile, zeilinger1979magnetic}} \\
\bottomrule
\end{tabular}
\end{table}

\begin{figure}
    \centering
    \includegraphics[width=1\linewidth]{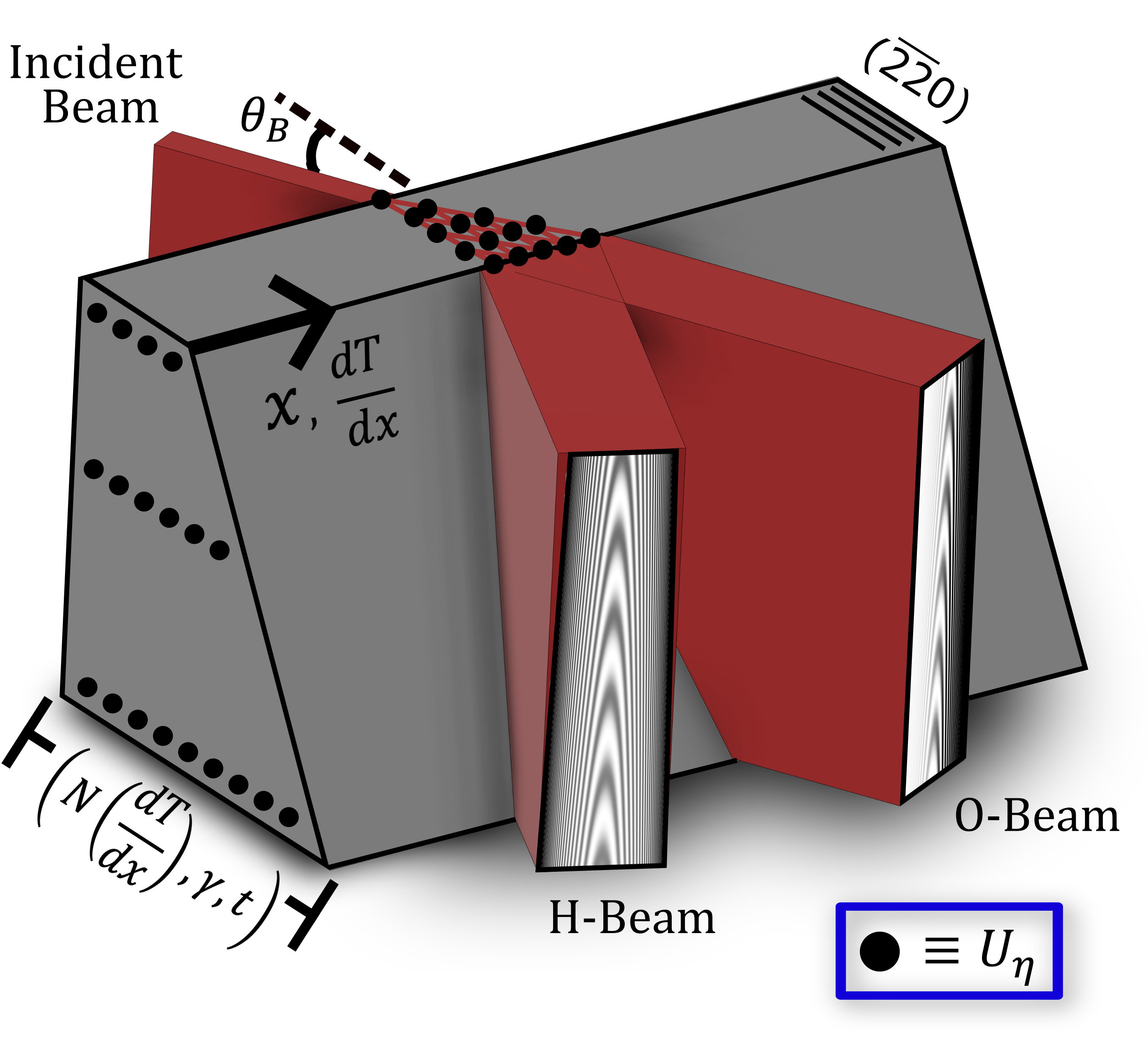}
    \caption{Schematic of the QI model implementation of a temperature gradient $dT/dx$ in a perfect silicon crystal wedge, with varying thickness $t$. A tall incident beam undergoes Bragg diffraction producing \Pend interference fringes in the O- and H-Beam observed at a 2D position sensitive detector. The temperature gradient is normal to the Bragg planes and modifies the number of simulation nodes $N$ with respect to the undeformed case (Eq.~\ref{eq:gamma}). The black circles represent the QI model nodes which apply the unitary Bragg operator (Eq.~\ref{U}).}
    \label{schematic}
\end{figure}

To observe this effect, consider a tall incident beam to a crystal wedge at the Bragg angle, whereby the crystal thickness is varying along the vertical extent of the beam (see Fig.~\ref{schematic}). This thickness variation reveals the \Pend interference patterns in both the transmitted and diffracted beams, observable with a position sensitive detector. Depending on the magnitude of the \Pend length, a precise number of intensity maxima will be visible within the wedge thickness range:
\begin{equation}
    t = \overline{n}_0 \Delta_H,
    \label{eq:fringe}
\end{equation}
where $\overline{n}_0$ is the interference order (or number of maxima). The \Pend oscillations shown at the 2D position sensitive detectors in Fig.~\ref{schematic} represent the perfect crystal case where $dT/dx = 0$. 

As shown in Ref.~\cite{hart}, computing the phase difference between interfering paths within the crystal reveals which interference order $\overline{n}$ will be observed as a function of the deformation $\frac{dT}{dx}$:
\begin{equation}
    \overline{n} = \frac{1}{2} \overline{n}_0 \left( \sqrt{1 + \overline{p}^2} + \frac{\arcsinh{(\overline{p})}}{\overline{p}} \right),
    \label{eq:hart}
\end{equation}
where
\begin{equation}
    \overline{p} = \frac{\alpha t \tan{\theta_B}}{|C||\chi_h|}  \frac{dT}{dx},
    \label{eq:p}
\end{equation}
$\alpha$ is the thermal expansion coefficient, $\frac{dT}{dx}$ is the linear temperature gradient, and $\chi_H = (-\lambda^2 F_H) / (\pi V)$.  Notably, Eq.~\ref{eq:hart} describes the observed fringe order $\overline{n}$ for a certain crystal thickness $t$ and deformation $\frac{dT}{dx}$, relative to the fringe order with no deformation $\overline{n}_0$. The fringe order is increasing as the magnitude of the deformation increases, implying that the fringe spacing must decrease. 

\begin{figure}
    \centering
    \includegraphics[width=\linewidth]{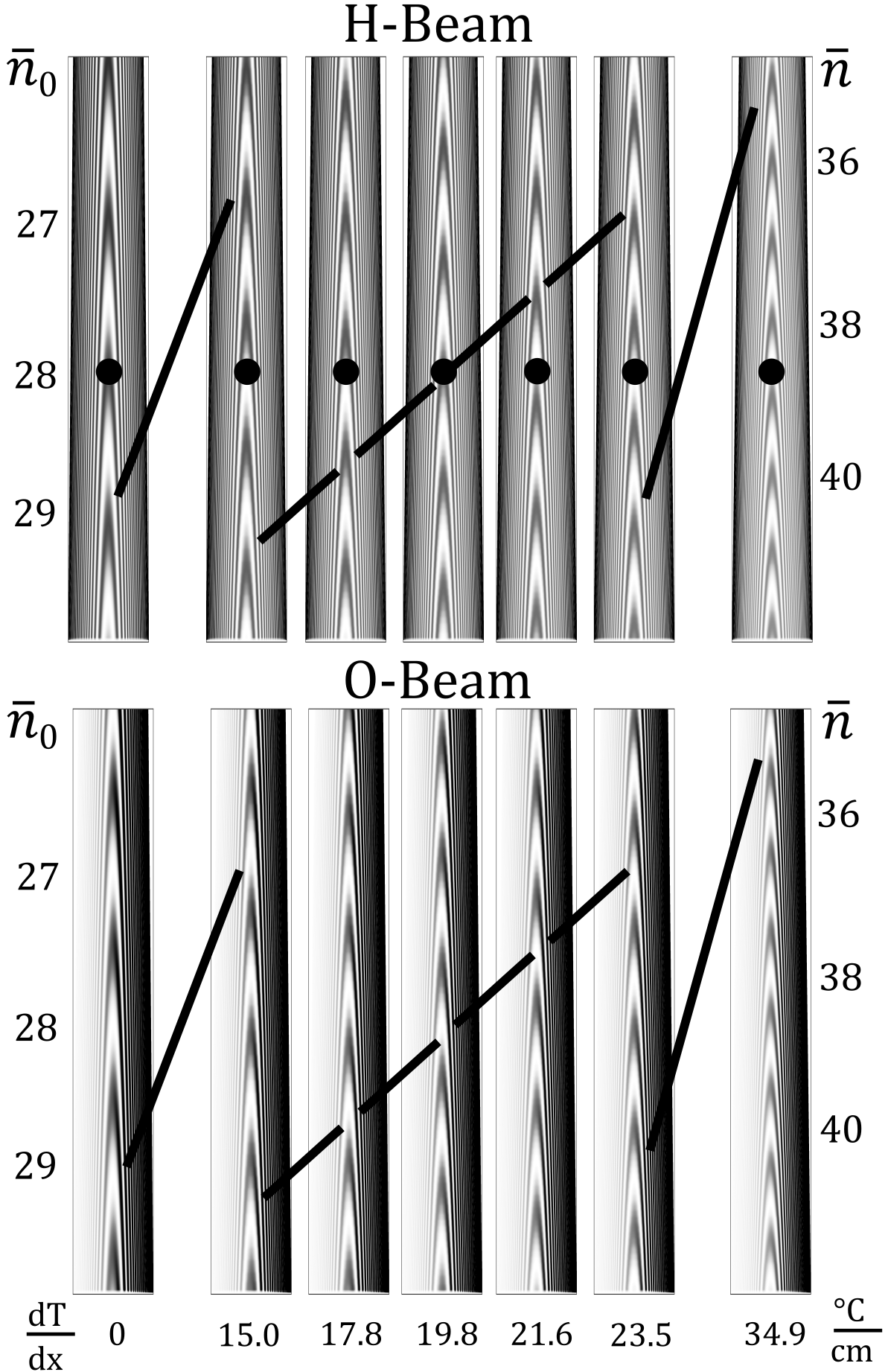}  
    \caption{Simulated O- and H-Beam interference patterns for $\overline{22}0$~Ag~K$\alpha_1$ incident on a crystal wedge subject to linear temperature gradients (see Fig.~\ref{schematic}) with magnitudes ($0 \leq \frac{dT}{dx} \leq 34.9~\degree$C/cm). The interference order in the ideal (deformed) case is indicated by $\overline{n}_0$ ($\overline{n}$). The solid black lines across the interference patterns connect identical interference orders: a consequence of the fringe spacing decreasing as the deformation increases. From Eq.~\ref{eq:equivalence},~\ref{eq:gamma}, we set $\gamma = \pi/100$ and vary $N$ according to the temperature gradient. The horizontal line of black circles in the H-beam intensity profiles indicate where the fringe order ratio $\overline{n}/\overline{n}_0$ is computed for Fig.~\ref{fig:curve}. The intensity profiles show qualitative agreement with those in Ref.~\cite{hart}; a quantitative comparison is demonstrated in Fig.~\ref{fig:curve}.}
    \label{fig:sim_temp}
\end{figure}

We now apply the QI model to simulate this temperature gradient. Using Eqs.~\ref{eq:equivalence}, ~\ref{eq:fringe}, and~\ref{eq:hart}, we obtain:
\begin{equation}
    \frac{N}{N_0} \approx \frac{\overline{n}}{\overline{n}_0},
    \label{eq:gamma}
\end{equation}
where $N_0$ ($N$) defines the depth of the quantum random walk for a crystal without (with) deformation. Consistent with the physical effect of elastic deformation, variation of $N$ in Eq.~\ref{eq:gamma} does not correspond to a change in the number of atomic sites, but instead represents a change in the lattice spacing. Shown in Fig.~\ref{fig:sim_temp} are the simulated \Pend oscillations for the diffracted (H-Beam) and transmitted (O-Beam) intensity profiles. The simulation considers incident $\overline{22}0$~Ag~K$\alpha_1$ X-rays to a crystal wedge ranging in thickness from about 1.20 mm to 1.45 mm~\cite{hart}. The interference order in the ideal (deformed) case is indicated by $\overline{n}_0$ ($\overline{n}$) and solid black lines connect identical interference orders across the different magnitude temperature gradients. The leftmost intensity profiles in the O- and H-Beam represent the perfect crystal case where $dT/dx = 0$. As the temperature gradient is introduced, we observe the movement of the fringes and the fringe spacing decreasing, in qualitative agreement with experiment~\cite{hart}.

\begin{figure}
    \centering
    \includegraphics[width=\linewidth]{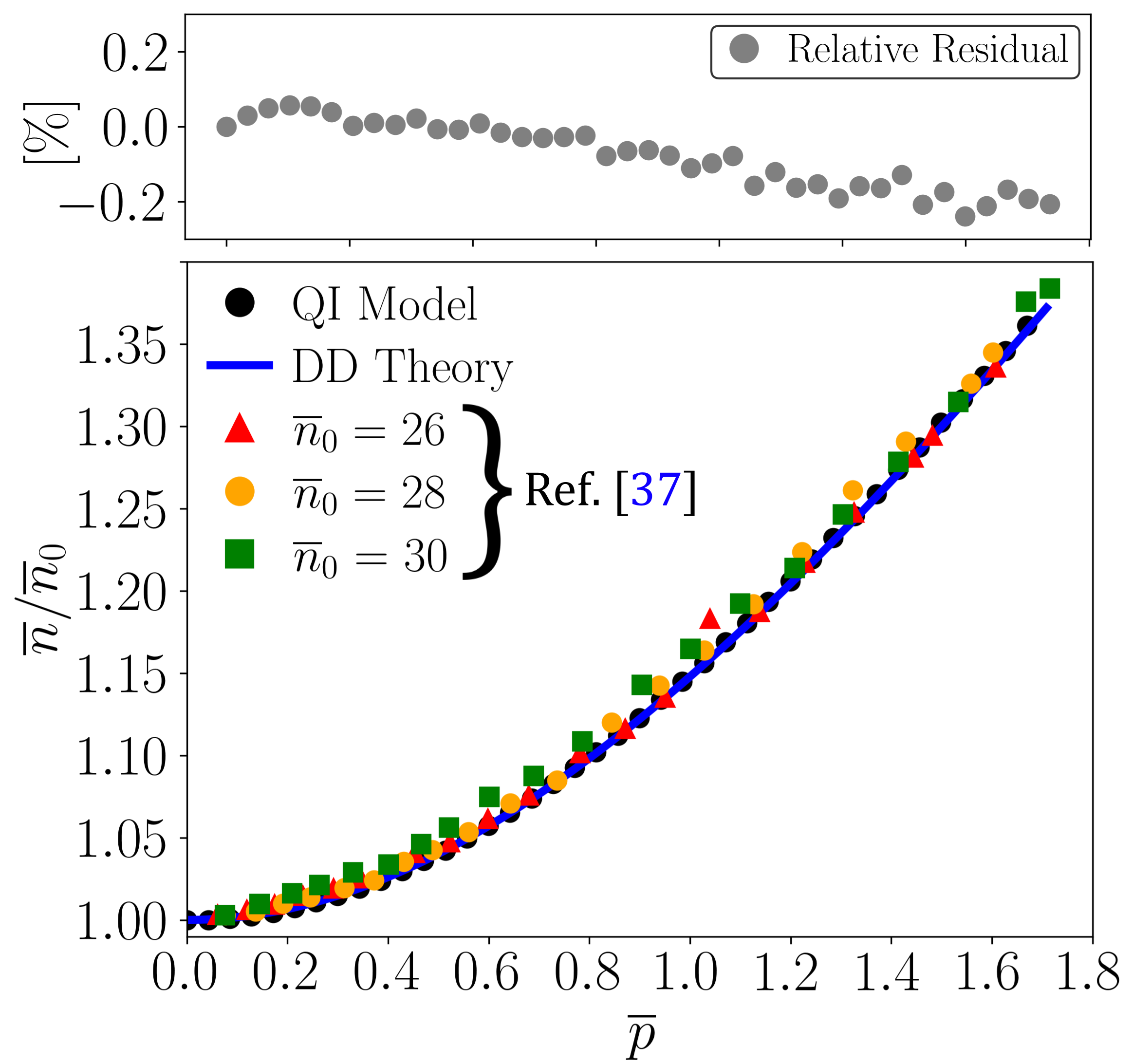}
    \caption{The measured fringe order ratio $\overline{n}/\overline{n}_0$ for $\overline{22}0~Ag~K\alpha_1$, at various observation points $\overline{n}_0$, as a function of temperature gradient $dT/dx$ (Eq.~\ref{eq:p}) from Ref~\cite{hart}, as well as the prediction from DD theory given by Eq.~\ref{eq:hart}, and the QI model simulation (from Fig.~\ref{fig:sim_temp}). The relative residual percentage between the QI model results and the DD theoretical curve is shown above the data: $(\text{QI} - \text{DD}) / \text{DD}$.}
    \label{fig:curve}
\end{figure}

To quantitatively compare the measurement results to experiment and DD theory (Eq.~\ref{eq:hart}), we choose fixed points of observation in the section patterns (i.e $\overline{n}_0 = 26$, $28$, $30$) and then interpolate between interference orders to calculate $\overline{n}$ as the crystal deformation increases (as done in Ref.~\cite{hart}). The horizontal line of black circles in Fig.~\ref{fig:sim_temp} demonstrates where $\overline{n}$ is computed at the fixed observation point $\overline{n}_0 = 28$. Shown in Fig.~\ref{fig:curve} is the measured phase ratio $\overline{n}/\overline{n}_0$ versus the magnitude of the temperature gradient (Eq.~\ref{eq:p}), as well as the prediction from DD theory (Eq.~\ref{eq:hart}) and the QI model simulation results. 

As indicated by the relative residual, the QI model and DD theory agree within $0.2~\%$ over the full temperature range considered. The observed downward trend reveals a systematic deviation, with the QI model predicting a $\sim 0.2~\%$ larger $\overline{n}/\overline{n}_0$ at the highest gradient ($dT/dx = 34.9~^\circ$C/cm). The small oscillations in the residual are an artifact of the simulation resolution, defined by setting $\gamma = \pi/100$. We compute the RMS error for both the QI model and DD theory compared to the $\overline{n}_0 = 28$ experimental data. While both models agree with the data at the $10^{-3}$ level, the QI model provides a slightly improved fit, with an RMS error approximately 14\% lower than that of DD theory. Overall, this agreement demonstrates that the QI model provides a simple yet quantitatively accurate framework for capturing the modification of \Pend oscillations under a linear temperature gradient.

\subsection{Fankuchen Cut Crystals}
In practice, crystal surfaces are not perfectly parallel (in Bragg geometry) or perpendicular (in Laue geometry) to the Bragg planes, but deviate within certain tolerances depending on the application ($\sim$ mrad). Such deviations can arise from miscuts or polishing tolerances and must be accounted for to achieve quantitative agreement with experimental results~\cite{nahman2022generalizing, gentile, sterbinsky2021simple}.

\begin{figure}
    \centering
    \includegraphics[width=\linewidth]{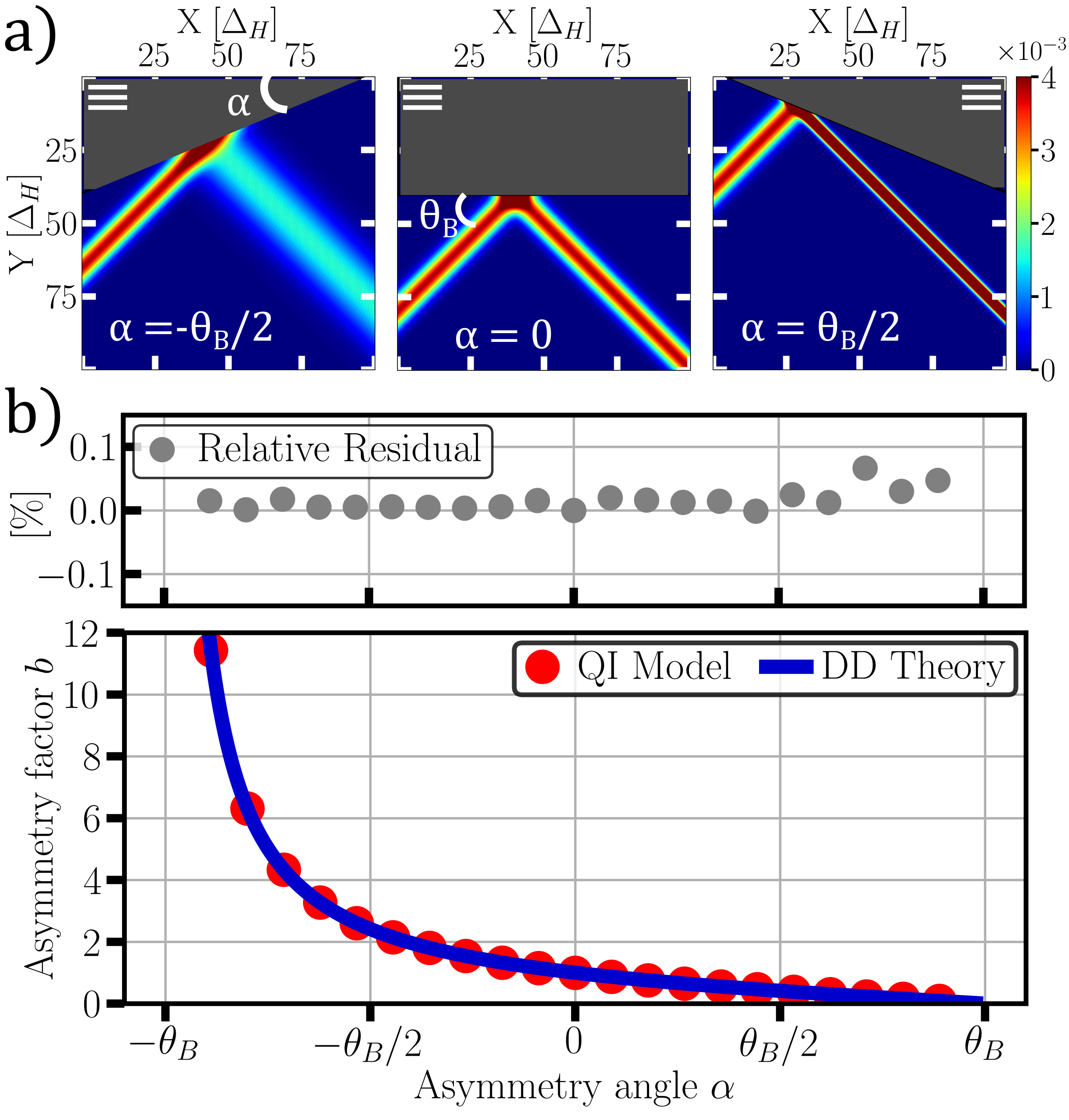}
    \caption{(a) QI model simulations of Bragg diffraction at $\theta_B = 45\degree$ from crystals with asymmetry angles $\alpha = -\theta_B/2$, $\alpha = 0$, and $\alpha = \theta_B/2$. The input Gaussian width is $w_{in} = 4\Delta_H$ and $\gamma = \pi/100$ from Eq.~\ref{eq:equivalence}. (b) Comparison of the QI model and DD theory for $w_{out}/w_{in} = |b|$, where $\theta_B = 45\degree$ and $\alpha$ is varied from $-\theta_B$ to $\theta_B$. The relative residual (QI - DD) / DD demonstrates agreement at the $0.1~\%$ level.}
    \label{fankuchen}
\end{figure}

Intentional angular offsets can also be introduced to tailor beam properties, as proposed by Fankuchen in the design of condensing X-ray monochromators~\cite{fankuchen1937condensing}. In this approach, the crystal surface is cut at an angle $\alpha$ relative to the Bragg planes (see Fig.~\ref{fankuchen}a), placing it within the broader framework of asymmetric Bragg diffraction~\cite{afanas1992asymmetric, eichhorn1983study, brauer1995perfect}. These angled crystal geometries have been studied for both X-rays and neutrons~\cite{evans1948aparallel, albertini1977evidence, wagh2011plain}.

Such controlled asymmetry is particularly valuable for neutron optics, where shaping or focusing beams without losses is challenging. In contrast to aperture-based or fresnel zone plate-based methods, asymmetric crystal geometries enable phase-space transformations that preserve beam intensity and enable beam expansion/condensing.

From DD theory, the crystal asymmetry factor $b$ is given by:
\begin{equation}
    b = \frac{\sin(\theta_B - \alpha)}{\sin(\theta_B + \alpha)},
\end{equation}
transforming the beam width as:
\begin{equation}
    w_{out} = |b|w_{in},
    \label{asym}
\end{equation}
where $w_{in}$ $(w_{out})$ are the input (output) beam widths from the crystal. In accordance with Liouville's theorem, the angular divergence scales as $1/|b|$, such that a compressed beam becomes more divergent. Multiple asymmetric crystals can be arranged in sequence to enable enhanced control over beam shaping and collimation.

In Fig.~\ref{fankuchen}, we model Fankuchen-cut crystals using the QI framework and obtain excellent agreement with DD theory. Analogous to the double Bragg blade cavity geometry considered in previous work~\cite{neutron_cav}, the crystal geometry is implemented by introducing free-space nodes,
\begin{equation}
    U_{\mathrm{free}} = U_i(\gamma = 0),
    \label{Ufree}
\end{equation}
which allow the crystal surface to be oriented according to the asymmetry angle $\alpha$.

We simulate an incident Gaussian beam with RMS beam width $w_{\mathrm{in}} = 4\,\Delta_H$ impinging on a crystal with Bragg angle $\theta_B = 45^\circ$, while varying the asymmetry angle over the range $-\theta_B < \alpha < \theta_B$. The output beam is fit to a Gaussian profile to extract $w_{\mathrm{out}}$. The plotted relative residual between the QI model and Eq.~\ref{asym} shows agreement within $0.1~\%$. As expected, taking the Fourier transform of the reflected beam profiles confirms that the angular divergence scales as $1/|b|$.

Notably, the QI model reveals that the DD interference responsible for reshaping the beam occurs directly at the crystal surface, with minimal penetration into the crystal. Identity nodes (Eq.~\ref{Ufree}) represent free-space propagation, while the standard Bragg diffraction operators (Eq.~\ref{U}) describe the crystal, as indicated by the grey regions in the figure. This capability to model imperfect or arbitrarily shaped crystals is critical for characterizing beam size and divergence in the design and analysis of neutron and X-ray crystal optics.

\subsection{Internal Talbot Effect}
\label{talbot_section}
The Talbot effect, in which a periodic object produces self-images at specific propagation distances, has found broad applicability across a range of fields~\cite{talbot1836facts, wen2013talbot}. The corresponding self-imaging distance, or Talbot distance $z_T$, is given by
\begin{equation}
    z_T = \frac{\lambda}{1 - \sqrt{1 - \frac{\lambda^2}{D^2}}},
\end{equation}
where $\lambda$ is the wavelength and $D$ is the periodicity of the object or grating. In the limit $\lambda \ll D$, this reduces to $z_T \approx 2D^2/\lambda$. Recently, the Talbot effect has been theoretically investigated by Balyan as an internal crystal phenomenon within DD theory~\cite{balyan2019a, balyan2019b, balyan2019c, balyan2020spherical, balyan2020x, balyan2021x}. In this context, the effect modulates the \Pend oscillations, generating interference ``carpets'' in the transmitted and reflected beams when an incident neutron or X-ray plane wave is structured by a periodic grating. Here, we show that this internal interference modulation is captured exactly within the QI framework.

The corresponding Talbot distance in a crystal is given by
\begin{equation}
    z_{td}^{cr} = \frac{2\Delta_H}{\sqrt{1 + 4 \left(\frac{\Delta_H \tan(\theta_B)}{D} \right)} - 1}.
    \label{talbot}
\end{equation}
If the condition $\Delta_H \tan(\theta_B) \ll D$ is satisfied, then Eq.~\ref{talbot} can be approximated as $z_{td}^{cr} \approx D^2/(\Delta_H \tan^2(\theta_B))$. Similarities with the optical case are clear, with the characteristic \Pend length scale replacing wavelength as the parameter of interest.

\begin{figure*}
    \centering
    \includegraphics[width=\linewidth]{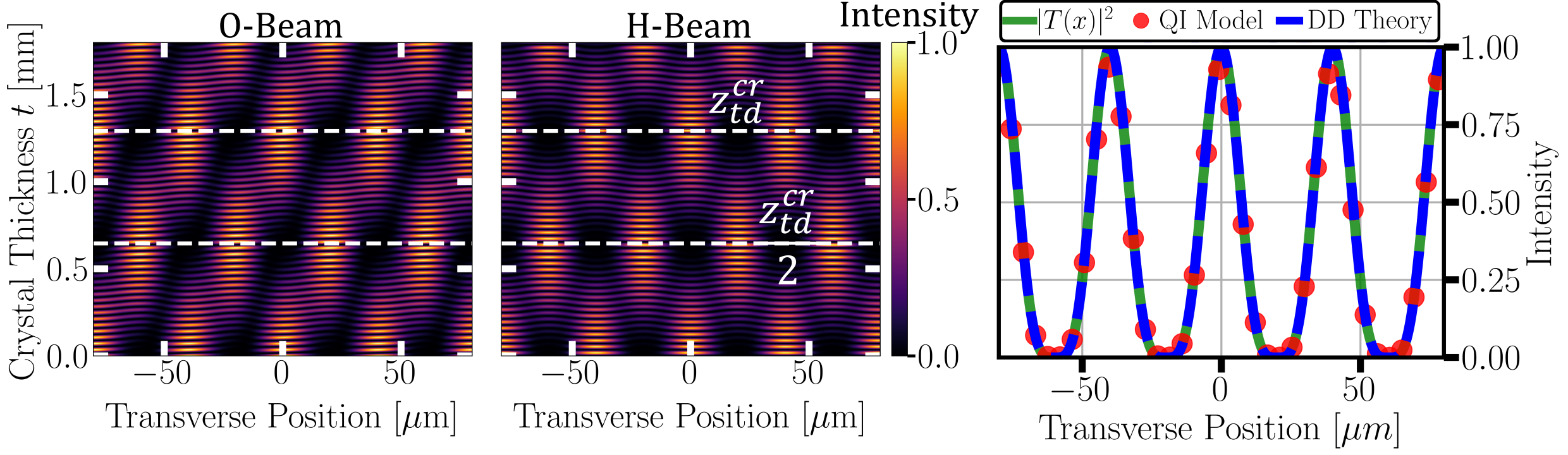}
    \caption{Simulated DD Talbot carpets in both the O- and H-Beam in the Laue geometry for a periodic amplitude grating with $D = 40~\mu$m. The Talbot distance is $z_{td}^{cr}\approx 1.3$~mm. As shown in the rightmost plot, the QI model self-image (transverse intensity at $t = z_{td}^{cr}$ in the H-Beam) matches the incident grating periodicity and DD theory result. These simulations neglect the effects of absorption.}
    \label{fig:talbot}
\end{figure*}

With plane-wave DD validated within the QI model (see Sec.~\ref{qi_sec}), we can incorporate the internal DD Talbot effect~\cite{balyan2019a}. We modulate an incident plane wave with a periodic amplitude grating of the form
\begin{equation}
T(x) = \frac{1 + \cos(2\pi x / D)}{2},
\end{equation}
with period $D = 40~\mu\mathrm{m}$. The resulting simulated Talbot carpets in both the O- and H-Beam are shown in Fig.~\ref{fig:talbot}. We consider (220) silicon with MoK$\alpha$ radiation ($\lambda = 0.71~\mathrm{\AA}$) and $\sigma$-polarization, yielding a \Pend length of $36.6~\mu\mathrm{m}$ and a Talbot distance of $z_{td}^{cr} \approx 1.3~\mathrm{mm}$. The O-Beam carpet appears tilted due to asymmetry in the transmitted DD wavefield~\cite{balyan2019b}.

At the Talbot distance, the reflected intensity reproduces the incident grating $T(x)$, confirming the expected self-imaging behavior and agreeing with DD predictions~\cite{balyan2019b}. At half the Talbot distance, $z_{td}^{cr}/2$, the carpet pattern exhibits the characteristic $D/2$ lateral shift.

Complementary to the results for a linear temperature gradient (see Fig.~\ref{fig:sim_temp}), this internal Talbot effect can be experimentally verified using wedged crystals to produce interference fringes on a 2D position-sensitive detector, as proposed in Ref.~\cite{balyan2019a}. Notably, the Talbot distance in Fig.~\ref{fig:talbot} is approximately 35 times larger than the \Pend period. This magnified spatial scale is ideal for utilizing neutron imaging detectors, where typical pixel sizes ($\sim 50~\mu\mathrm{m}$) are well-suited to resolve the Talbot structure, but not for the much finer \Pend oscillations. As a result, the Talbot effect provides a practical route for imaging internal DD behavior, whereby we propose that the \Pend length can be retrievable from the measured carpet periodicity.

\section{Discussion and Conclusion}
In this work, we have established a comprehensive quantum random walk model capturing known dynamical diffraction phenomena. While traditional analytical DD approaches become impractical when confronted with complex crystal geometries, deformations, and imperfections, our quantum random walk framework naturally accommodates these real-world experimental environments. By successfully integrating the coherent plane-wave and spherical-wave regimes into a single model, this work provides an accurate and highly accessible toolkit for simulating both macroscopic crystal diffraction and complex internal crystal effects. 

Rather than merely reproducing individual diffraction effects, our framework demonstrates a robust capacity to naturally implement common crystal deformations and arbitrary crystal geometries. We validated this by accurately capturing the modified \Pend oscillations induced by linear temperature gradients, and by demonstrating the tunable beam transformations inherent to asymmetric Fankuchen-cut crystals. Furthermore, we utilized the plane-wave formulation to simulate the internal crystal DD Talbot effect. Notably, since this effect produces spatial intensity modulations on a scale resolvable by modern neutron imaging detectors, our model serves as an ideal predictive tool for the experimental design and verification of these internal interference carpets, offering a practical route to directly extract the orders of magnitude smaller \Pend length.

Looking forward, the natural flexibility of the unified quantum model positions it as an essential computational foundation for the development of next-generation perfect crystal interferometers and advanced neutron optics. As outlined in Table~\ref{table}, the framework is readily extensible to more complex scenarios. For example, intentionally bent crystals~\cite{rekveldt1987bent, mikula2011observation, guigay2022x} are widely used to create focusing monochromators and to probe fundamental neutron–matter interactions, such as with neutron whispering gallery states~\cite{nesvizhevsky2010neutron}. Analogous to the misset plane-wave approach shown in Fig.~\ref{fig:plane} and Eq.~\ref{plane}, these effects may be incorporated into the QI framework through position-dependent phase gradients across the node-discretized crystal. Additionally, magnetic fields and spin–orbit interactions are essential extensions for accurately describing neutron spin evolution both inside and outside the crystal~\cite{zeilinger1979magnetic, gentile}. These effects can be included by expanding the simulation space to apply a $2 \times 2$ unitary spin rotation operator, $U_{\text{spin}}$, at each node, analogous to the Bragg operator (Eq.~\ref{U}), but with the spin rotation angle arising from external magnetic fields or spin–orbit interactions. 

Ultimately, by providing a precise, versatile, and unified alternative to standard DD theory, this framework will be instrumental in characterizing experimental diffraction experiments, optimizing large-area interferometers, and enabling high-precision measurements of fundamental interactions.

\section{Acknowledgments}
This work was supported by the Canadian Excellence Research Chairs (CERC) program, the Natural Sciences and Engineering Research Council of Canada (NSERC) Discovery program, the NSERC Canada Graduate Scholarships programs (CGS-M and PGS-D), the Collaborative Research and Training Experience (CREATE) program, the Canada  First  Research  Excellence  Fund  (CFREF), and the US Department of Energy, Office of Nuclear Physics, under Interagency Agreement 89243019SSC000025.

\bibliographystyle{ieeetr}
\bibliography{mybib.bib}
    
\end{document}